\documentclass[10pt,conference]{IEEEtran}
\usepackage[T1]{fontenc}
\usepackage{cite}
\usepackage{amsmath}
\usepackage{booktabs}
\usepackage{graphicx}
\usepackage{xcolor}
\usepackage{url}
\usepackage[hidelinks]{hyperref}
\usepackage{pgfplots}
\usepackage{array}
\usepackage{tabularx}
\pgfplotsset{compat=1.17}
\usetikzlibrary{positioning,arrows.meta,fit}

\newcommand{\sys}{\textsc{Ledger}}
\newcommand{\hyb}{\textsc{Ledger+Ep}}

\begin{document}

\title{LEDGER: Auditable Agent Memory Under Change}


\author{
\IEEEauthorblockN{
Md Aidul Islam, Md Nasir Uddin Shuvo,\\
Muhammad Waseem, Pekka Abrahamsson
}
\IEEEauthorblockA{
\textit{Faculty of Information Technology and Communication Sciences}\\
\textit{Tampere University}\\
Tampere, Finland\\
\{mdaidul.islam, nasir.shuvo, muhammad.waseem, pekka.abrahamsson\}@tuni.fi
}
}

\maketitle
\pagestyle{plain}
\thispagestyle{plain}

\begin{abstract}
Agent memory systems are mostly evaluated on fixed histories, and only their final answers are scored. In real use, facts change, sources differ in reliability, and users ask the agent to forget things. We present \sys{}, a memory layer in which a language model proposes facts and explicit rules decide which value is current, which source to trust, and what to delete. Each fact keeps its source and the period in which it held. \hyb{} adds a search over raw conversation turns for questions the ledger does not cover, and scrubs these turns on deletion. We compare both against no memory, full-history replay, retrieval over raw transcripts, and Mem0, with the same answering model and prompt, on a synthetic multi-session coding-agent workload and a 210-question subset of LongMemEval, and we score both the evidence each memory sends to the model and the answer. On the coding workload, \hyb{} answers 94.3\% of questions correctly on five development seeds and 91.1\% on five held-out seeds, against 82.1\% and 83.3\% for full replay, the strongest baseline. When a deletion request only describes the fact, no deleted value reached our systems' evidence in any evaluated probe. On LongMemEval, \hyb{} has the same observed accuracy as full replay (74.8\%) with 1\% of its context. The main limitation is that the coding workload is synthetic and built by us; on LongMemEval, the differences between the strongest systems are within the confidence intervals.
\end{abstract}


\section{Introduction}

Coding agents work on the same repository across many sessions~\cite{joshi2025swe, rakotonirina-etal-2025-tools}. The repository and its documentation remain available, whereas conversational context from earlier sessions is not automatically carried forward not~\cite{ packer2023memgpt}. Without memory, the agent has to recover these facts again, or the user has to repeat them: which database a service uses, which branch it deploys from, who is on call. A persistent memory carries such facts from one session to the next.

Most of these facts change over time. They appear in user messages, tool output, documents, and code changes. A fact may be updated, restated, or contradicted by a stale document~\cite{hu2026evaluating},or by the agent's own earlier erroneous inference~\cite{zhang2023language}. Sometimes the user asks the agent to forget it~\cite{xie2026can}. A memory that returns an old value next to the current one, trusts a guess over the user, or brings
back a deleted value can carry that error into later
sessions~\cite{xiong2026memory,xie2026can}.

Two simple approaches are the standard baselines in recent memory benchmarks~\cite{wu2025longmemeval,hu2025memoryagentbench}. Full-history replay sends the whole past conversation with every question. Nothing is lost, but the context grows with the history, and old, wrong and deleted values reach the model with the current one. Retrieval over transcripts (transcript RAG) sends only the few past turns most similar to the question. The context stays small, but the search can miss the right turn or return an outdated one. In both cases, the model has to decide on every question which value is current. Dedicated memory systems~\cite{packer2023memgpt,chhikara2025mem0,rasmussen2025zep,xu2025amem} reduce what the model reads, but the model still decides when a stored fact is replaced or removed, and their described mechanisms do not generally provide explicit, user-configurable source-reliability rankings.

To the best of our knowledge, no prior benchmark jointly evaluates memory correctness, source trust, deletion compliance, and evidence-level provenance in multi-session coding or conversational agent settings. Existing benchmarks such as LoCoMo and LongMemEval evaluate long-term memory primarily through downstream question answering; LoCoMo additionally reports retrieval recall, but neither benchmark~\cite{maharana2024locomo,wu2025longmemeval} audits whether the evidence ultimately provided to the model reflects the current, trusted, and non-deleted state of the underlying facts.

The objective of this study is to test whether an agent memory can stay correct when facts change, when sources differ in reliability and when users ask to forget, while sending the model far less text than full replay. We ask three questions:
\begin{itemize}
    \item\emph{\textbf{RQ1:}} How well does the complete system, with LLM extraction and routing, keep stale, contaminated and deleted values out of the evidence it sends to the model?
    \item\emph{\textbf{RQ2:}} Does such a memory answer better than replay, transcript RAG and an existing memory system under the same answering model and prompt?
    \item\emph{\textbf{RQ3:}} As the history grows, does it stay correct, and how do its context, storage and model cost grow?
\end{itemize}
We move some decisions out of the model. In \sys{}, a language model still extracts candidate facts and routes each question to stored facts. Fixed rules then decide which value is current, which source wins a conflict, and what a deletion erases. \hyb{} adds a search over raw conversation turns for questions the ledger does not cover. Our contributions are:

\begin{enumerate}
    \item \emph{Memory design.} We propose \sys{}, a fact-centric memory system that maintains current values, validity intervals, provenance, and deletion state under five ingestion rules governing updates, conflicts, and source precedence. \hyb{} extends \sys{} by retaining raw turns, which are scrubbed upon deletion.

    \item \emph{Evidence-level evaluation.} We introduce an evaluation protocol that audits both the evidence packet supplied to a fixed answering model and the resulting answer, classifying retrieved errors as stale, conflicting, untrusted, or deleted. The protocol reports answer accuracy, retrieval correctness, token use, cost, latency, and storage overhead on a synthetic coding-agent workload with controlled updates, conflicts, deletions, and a simulated 540-day replay.

    \item \emph{Controlled comparison.} We compare \sys{} and \hyb{} with no memory, full replay, transcript RAG, and Mem0 under identical model, prompt, and token-budget conditions. Evaluations cover development and held-out coding-workload seeds and a 210-question LongMemEval subset.
\end{enumerate}

Section.~\ref{sec:coverage} states which challenge requirements we cover and which, such as a downstream agent task, we do not.

\section{Problem Definition and Scope}
\label{sec:problem}

\subsection{Problem setting}

An agent works with one or more users over many sessions. Each session
produces a stream of \emph{events}: user and assistant messages, tool calls, tool results, documents and code changes. Every event has an id, a user, a session and a timestamp. Our replay uses these timestamps as a logical clock: sessions are ingested in timestamp order, events with the same timestamp keep their stream order, and an event counts as ingested at its timestamp. Some events state \emph{facts}: a property of an entity whose value can change, such as the database a service runs on. A memory reads these events. When the agent asks a question, the memory returns a small amount of evidence for an answering model, or abstains.

Facts have two kinds of time. \emph{Valid time} is the period in which a value was true. \emph{Recording time} is when the memory learned it; on the replay clock, this is the timestamp of the event that stated it. A question is asked at a time $t$. It can ask about now, or about an earlier time $t'$ (\emph{as of} $t'$). The correct answer is the value that was valid at $t'$ (or at $t$), based only on what was recorded up to $t$. \sys{} uses a restricted form of this setting, in which the two times are the same (A4).

When a value changes, the old value is not removed: its valid time ends at the change, so it leaves current answers but still answers questions as of an earlier date. Deletion removes a value from both.

\subsection{Supported operations}

Every system, ours and the baselines, is run through the same evaluation adapter. The adapter gives every system the same input, but not every system can act on it.
\begin{itemize}
\item \texttt{ingest}: add one session's events, in time order.
\item \texttt{retrieve}: given a question, the user, the time it is asked and
  optionally an as-of time, return evidence within a token budget, or
  abstain.
\item \texttt{delete}: forget a fact at the user's request. The request
  carries a text description and, when known, the ids of the events that
  stated the fact. We also test requests that carry only the description.
\end{itemize}
The runner sends a deletion request right after the user's ``forget'' message, which is also ingested as a normal turn, so no system has to detect deletion in the conversation. The transcript baselines and our Mem0 adapter delete only by event id, so a description-only request removes nothing through their delete operation; Mem0 can still act on the ``forget'' turn through its ordinary update step. This describes the evaluated configuration, not what Mem0 could support with another integration. Only \sys{} routes the request to stored facts.

\subsection{Desired properties}

Table~\ref{tab:inv} lists seven properties that a correct memory should have. Each one rules out one type of error, and we use these types to label wrong values in the evaluation. The properties describe what we want from any memory. The last column shows under which assumptions \sys{} guarantees each one. These guarantees are per slot: given the right slot, the rules pick the right entry. Reaching the right slot is a separate step, routing (A2).

\begin{table}[t]
\caption{Desired properties of a memory, the error each rules out, and the
assumptions under which \sys{} guarantees it.}
\label{tab:inv}
\centering
\footnotesize
\begin{tabular}{@{}l p{0.50\columnwidth} l l@{}}
\toprule
 & Property & Error & \sys{} \\
\midrule
I1 & \emph{Deletion}: a fact recorded before the user asked to forget it is never returned again, also for questions about the past. The user may state it again later & resurrected & A1, A2 \\
I2 & \emph{Supersession}: once an accepted update changes a fact, the old value is no longer the current answer. An update is accepted if its source is at least as trusted as the current value & stale & A1, A3 \\
I3 & \emph{Temporal}: a question as of $t'$ returns the value valid at $t'$,
     not a later one & future & A1, A4 \\
I4 & \emph{Deduplication}: restating a fact or re-ingesting an event creates no new value entry; a restatement may add provenance & -- & A1 \\
I5 & \emph{Contamination}: a less trusted source (a stale README, the
     agent's guess) does not override the user & contaminated & A1, A3 \\
I6 & \emph{Abstention}: when nothing relevant is known, the memory says so
     & no abstention & A2 \\
I7 & \emph{Isolation}: one user's memory never reaches another user & -- &
     user id \\
\bottomrule
\end{tabular}
\end{table}

\subsection{Assumptions and scope}
\label{sec:assumptions}

\sys{} guarantees the properties only under four assumptions:
\begin{description}[\IEEEsetlabelwidth{A4}]
\item[A1] \emph{Correct extraction.} For every event that states a fact,
  the extractor returns a claim with the right value and the same slot key
  each time the fact is mentioned. It returns no claim for text that does
  not assert a value, such as a question, a plan or a negation.
\item[A2] \emph{Correct routing.} Each question and deletion request is
  mapped to all the slots it is about, and to no other. A deletion request
  whose source events were extracted does not need routing.
\item[A3] \emph{Trust order.} The source    ranking matches how reliable sources are. A document that is newer and correct but conflicts with the user is disputed, not accepted.
\item[A4] \emph{Valid time starts when a value is stated.} We do not read
  dates written in the text. Each entry stores only a start time, the
  timestamp of the event that stated the value, and an end time, the
  timestamp of the event that replaced it; on the replay clock, the
  recording time equals the start time. A change reported late, such as ``we migrated last Tuesday'', is therefore dated to the message that reports it.
\end{description}
A1 and A2 are idealised conditions. Our extractor and router are LLM calls and do not always meet them; when they fail, the properties that depend on them can fail too, and Sec.~\ref{sec:fails} counts how often. I7 needs only that the caller passes the right user id and that every read and deletion goes through the interface, because all state is kept per user.

\hyb{} keeps I7, and I1 with one open case; the other properties hold for its ledger facts but not for the raw turns it adds. We study facts with one current value each, in English text, from coding-agent sessions and from LongMemEval chats.
\section{LEDGER Architecture and Mechanisms}
\label{sec:design}

\subsection{Architecture overview}
\label{sec:hybrid}

\begin{figure*}[t]
\centering
\includegraphics[width=\linewidth]{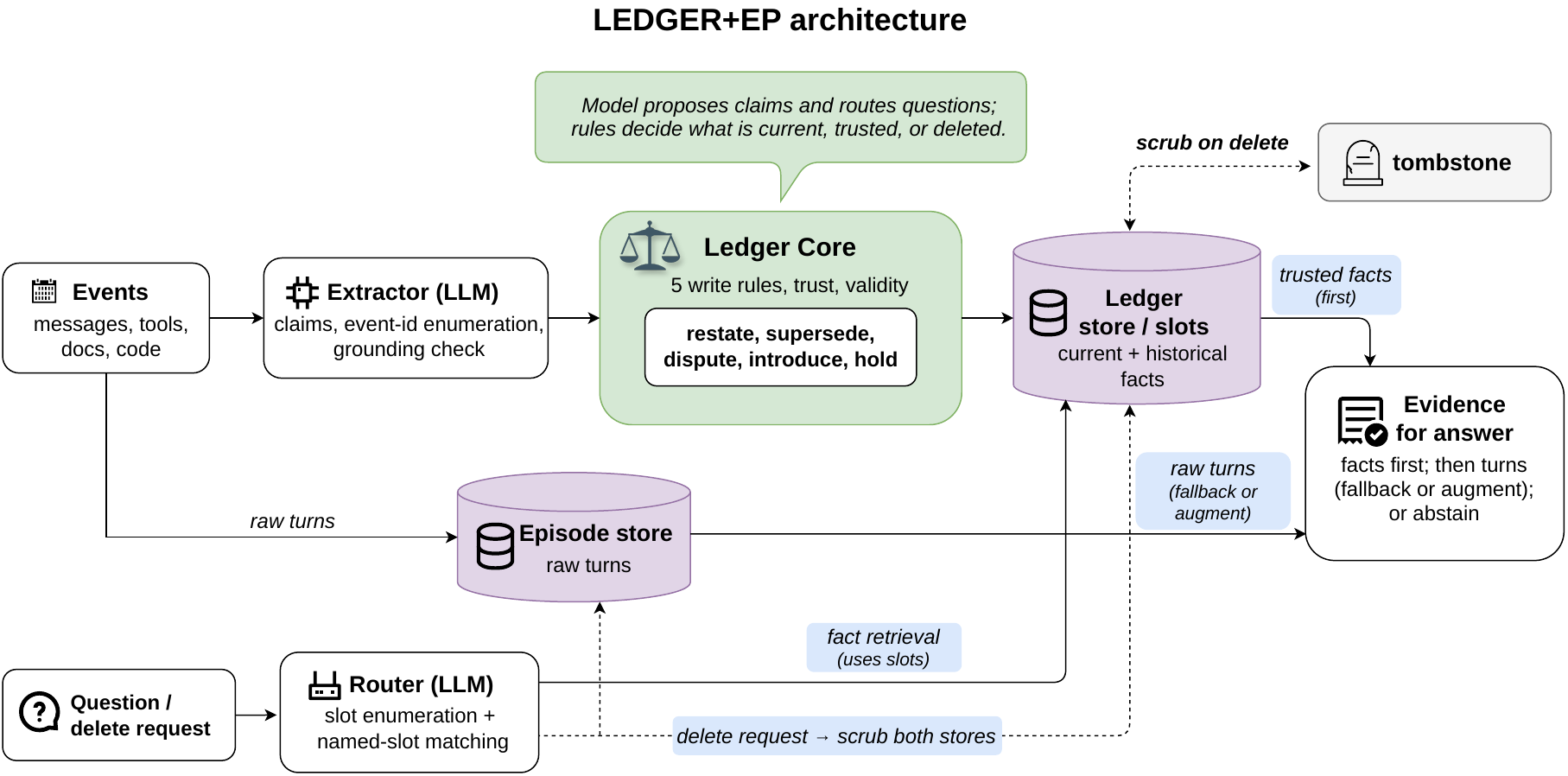}
\caption{Components of \sys{} and \hyb{}. \sys{} uses the extractor, the
router and the ledger core; \hyb{} adds the episode store. The model
proposes claims and routes questions; the ledger core alone decides what
is current, trusted or deleted, and the episode store is scrubbed on every
deletion.}
\label{fig:arch}
\end{figure*}

Figure~\ref{fig:arch} shows the components. An \emph{extractor} reads each ingested session and proposes \emph{claims}: this event states that this property has this value. The \emph{ledger core} applies fixed write rules to each claim, and it alone decides what is current, trusted or deleted. A \emph{router} maps each question or deletion request to the slots the ledger knows. In our main setup the extractor and the router are calls to a language model. A rule-based extractor serves as a model-free floor.

We evaluate three variants:
\begin{itemize}
\item \sys{}: the ledger alone. It returns one fact line per matched slot, or
  abstains.
\item \hyb{} \emph{fallback}: adds an \emph{episode store} of raw turns. It
  returns the ledger's facts whenever the ledger has any, and BM25-ranked
  turns~\cite{robertson2009bm25} only when it has none.
\item \hyb{} \emph{augment}: always returns the ledger's facts first, then
  fills the rest of the token budget with turns.
\end{itemize}

The episode store exists because the ledger models only facts with one current value. Some questions are not about such facts: what the assistant recommended, a one-off detail in a long chat, a count across sessions, or a fact whose first statement the extractor missed. Raw turns are not filtered by the rules, so they can carry an old or less trusted value; Sec.~\ref{sec:retrieval} states when each mode adds them.

\subsection{Data model and time}
\label{sec:ledger}

Memory is partitioned by user. A \emph{slot} is a property of an entity, keyed \texttt{entity/attribute} (e.g.\ \texttt{payments-service/database}). Each value a slot has held is an \emph{entry} with a validity interval $[\mathit{from}, \mathit{to})$~\cite{snodgrass1999temporal}. The interval starts at the timestamp of the event that states the value (A4) and ends when a newer value supersedes it. A question asked at $t$ as of $t'$ returns the entry that was valid at $t'$ and recorded by $t$. An entry also keeps the source kind and trust of the event that set it, and the ids of every event that stated it. When a more trusted source restates the value, the entry takes the higher trust, so a later conflicting document is disputed.

Besides entries, the ledger keeps three kinds of records that are never
retrieved: \emph{disputed} values from less trusted sources, kept for audit; \emph{unconfirmed} values, held until a trusted source states a value; and \emph{tombstones}, which record that a slot was deleted but not its value.

\subsection{Ingestion and conflict resolution}
\label{sec:extraction}

\textbf{Extraction.} For each ingested session, the extractor makes one
structured-output call and returns claims
$\langle\mathit{event}, \mathit{slot}, \mathit{value}\rangle$. It sees the user's known slots and their current values, and is asked to reuse a known key for the same property. Keys are normalised only in form, so synonyms such as ``inventory'' and ``inventory-service'' become two slots. Claims are applied one at a time in the order of their events. Events already ingested are skipped by id (I4). The ledger core never reads free text. Two safeguards check the extractor's output before any rule applies:
\begin{enumerate}
\item \emph{Enumerated event ids.} The output schema lists the ids of the
  batch's events, so a claim can only cite an event that exists.
\item \emph{Grounding check.} A claim is kept only if its value occurs as
  text in its event, ignoring case and punctuation. This drops invented or copied values, but not a value in a negation or a sentence about another service.
\end{enumerate}
A third safeguard concerns deletion.

\textbf{Write rules.} The core applies one of five rules to each claim, in
stream order:
\begin{description}[\IEEEsetlabelwidth{supersede}]
\item[restate] same value as the current entry, from any source:
  add the event to its provenance and keep the higher of the two trust
  levels; no new value entry (I4);
\item[supersede] different value from a source at least as trusted as the
  current entry's: close the current entry at the claim's time and open a
  new one (I2); the closed entry still answers historical questions (I3);
\item[dispute] different value from a less trusted source: kept for audit,
  never retrieved (I5);
\item[introduce] no current value, source trust $\geq 2$: open an entry;
\item[hold] no current value, source trust $<2$: keep the value as
  unconfirmed. If a trusted source later states the same value, it becomes
  an entry valid from the trusted statement; if it states a different value, the held value becomes a
  dispute.
\end{description}
Table~\ref{tab:example} traces the rules on one fact.

\begin{table}[t]
\caption{The write rules on one slot, payments-service/database.}
\label{tab:example}
\centering
\footnotesize
\setlength{\tabcolsep}{3pt}
\renewcommand{\arraystretch}{1.15}

\begin{tabularx}{\columnwidth}{@{}
  r
  >{\raggedright\arraybackslash}p{0.19\columnwidth}
  >{\raggedright\arraybackslash}p{0.23\columnwidth}
  >{\raggedright\arraybackslash}X
@{}}
\toprule
Day & Source (trust) & Claim & Rule and effect \\
\midrule
1 & user (3)
  & PostgreSQL 14
  & introduce: entry $[1,\infty)$ \\

3 & README (2)
  & Postgres 12
  & dispute: kept for audit only \\

5 & user (3)
  & PostgreSQL 14
  & restate: event added to provenance \\

9 & migrate.log (3)
  & PostgreSQL 16
  & supersede: $[1,9)$ closed; $[9,\infty)$ opened \\

12 & user (3)
   & ``forget the database''
   & delete: both entries, the dispute and the events
     erased; tombstone retained \\
\midrule
\multicolumn{4}{@{}p{\columnwidth}@{}}{%
  \raggedright
  Asked on day 7: PostgreSQL 14;
  on day 10 as of day 6: PostgreSQL 14;
  on day 10: PostgreSQL 16;
  on day 13: abstain,
  ``deleted at the user's request''.
} \\
\bottomrule
\end{tabularx}
\end{table}

\textbf{Source trust.} Table~\ref{tab:trust} collects the trust policy. Each event's kind sets its trust. Documents can introduce a value, because they are often a fact's only first source, but cannot override the user, a tool result or a code change. The assistant's messages and tool calls can never set or change a value; \hyb{} therefore does not send them to the extractor, but keeps them in the episode store.

The ranking rests on assumption A3. We assume
the user knows their own setup; tool output and code changes observe the
system directly (a migration log, a diff); documents are often right but can
be stale; and the assistant's messages and tool calls are guesses or
intentions, not observations. When this assumption is wrong, the policy fails
in a known way: a newer and correct document cannot override an older user
statement, and a wrong tool result overrides the user.

\begin{table}[t]
\caption{Source trust policy: what each source may do in the ledger, and
whether it is sent to the extractor in each variant.}
\label{tab:trust}
\centering
\footnotesize
\setlength{\tabcolsep}{3pt}
\renewcommand{\arraystretch}{1.15}
\begin{tabularx}{\columnwidth}{@{}
  >{\raggedright\arraybackslash}X
  c
  >{\raggedright\arraybackslash}p{0.13\columnwidth}
  >{\raggedright\arraybackslash}p{0.21\columnwidth}
  >{\raggedright\arraybackslash}p{0.20\columnwidth}
@{}}
\toprule
Source & Trust & Introduce & Update &
Extracted\newline (\sys{} / \hyb{}) \\
\midrule
User message, tool result, code change
& 3 & yes & any value & yes / yes \\

Document
& 2 & yes & values of trust $\leq 2$ & yes / yes \\

Assistant message, tool call
& 1 & no (held) & never (disputed) & yes / no \\
\bottomrule
\end{tabularx}
\end{table}

\subsection{Retrieval and abstention}
\label{sec:retrieval}

The router maps the question to slot keys chosen from a list of the user's known slots, so it cannot invent a slot. For each matched slot, the ledger returns the one entry valid at the question's as-of time and recorded by the time it is asked, as a single line with its validity and source. Matched slots without such an entry are left out. When none of them has one, the ledger abstains with a reason: ``no known slot matches'', ``deleted at the user's request'' or ``only an unconfirmed claim'' (I6). Under A1 and A2, the evidence therefore never contains a disputed, unconfirmed or deleted value.

In \hyb{}, the episode store ranks the user's stored turns with BM25 and adds the top five, each with its date and source kind, while they fit in the token budget left after the facts. Turns are not filtered by the as-of time. Fallback mode searches only when the ledger returns no fact; augment mode always adds turns after the facts. A question that matches only deleted or empty slots is refused in both modes. If it matches a deleted slot and a live one, augment mode still adds turns, so here only episode scrubbing keeps the deleted value out.

\subsection{Deletion}
\label{sec:deletion}

\textbf{Affected slots.} If the request carries source event ids, the affected slots are those with any entry, dispute or unconfirmed value from these events. An id cannot single out one fact of an event that states several, so all slots stated by these events are affected; in our coding workload each event states at most one fact. If there are no ids, or the extractor never turned these events into claims, the router maps the description to known slots. As a third safeguard, we also add every slot the description names outright. One routing call can return no slot, and a missed slot is a privacy failure.

\textbf{Ledger deletion.} For each affected slot, the ledger erases every entry, dispute and unconfirmed value recorded up to the request, past values included (I1). Each affected slot keeps a tombstone without its value. The extractor is never shown an erased stored record again; a value the user states after the request opens a new entry, while questions about earlier times still abstain.

\textbf{Episode scrubbing.} \hyb{} removes from the episode store every event behind an erased value, every turn that mentions an erased value, every turn up to the request that names a deleted slot outright, and every assistant or tool-call turn up to the request that names an entity the request is about. The last two rules cover what the ledger never extracted. Scrubbing more turns lowers the risk of exposure but can also remove useful details; we did not measure this loss of recall. Every removed event is blocked, so a replay after a crash cannot ingest it again.

\textbf{Guarantee and limits.} After a deletion, no value recorded before the request is returned again, for current or past questions, by \sys{} or by either mode of \hyb{}. This holds when the affected slots are found (A2) and the extractor captured the statements of the value (A1). One case remains open in \hyb{}: a trusted turn whose value the extractor missed, and which names neither the deleted slot nor any erased value, stays in the episode store and can still be returned. Deletion does not reach earlier snapshots, the response cache or run logs, which a deployment would also have to clear.

\subsection{Persistence and isolation}

Each store writes its full state, including tombstones and blocked ids, to one JSON file and can restore it; \hyb{} writes both stores to the same file. In our runs a snapshot is taken after every checkpoint; the file is not written atomically, and we did not test a crash during a write. All state is partitioned by user, and retrieval and deletion only touch the partition of the user id they are given (I7).
\section{Evaluation Design}
\label{sec:method}

\subsection{Coding-agent workload with planted ground truth}
\label{sec:workload}

A seeded generator produces multi-user, multi-session coding-agent timelines.
Each user owns several services; each service has 3--5 facts drawn from nine
kinds (database, test command, deploy branch, staging host, runtime, message
queue, CI provider, cloud region, on-call owner). A state machine writes each
fact's history as \emph{mutations}: introduce, restate, update, contradict
(from a document or the assistant only) and delete (from the user only).
Every mutation is rendered by a template of its source: user message, tool
output, code diff, document, or assistant message, alongside chit-chat,
tool calls and distractors that name a topic without a value. At checkpoints the generator asks probes of six types: \emph{current}, \emph{post-update}, \emph{historical} (with an as-of time), \emph{contamination}, \emph{post-delete} and \emph{abstention}. The complete generator settings (mutation weights, source mix, probe counts per type) are in the supplement (S1).

\textbf{Ground truth.} Expected answers come from a function of the workload code that reads only each mutation's kind and time: the answer is the last introduced or updated value at the target time, after the last deletion. It never calls a memory system and never reads event kinds or trust. A text check verifies that no event mentions a value it does not carry, and values are unique per user. We wrote the workload, the oracle and \sys{} together, and the workload encodes the same trust policy: contradictions come only from documents and the assistant. It therefore tests whether a memory follows this policy, not whether the policy is right (A3). Held-out seeds and LongMemEval are our checks against fitting the workload.

We use two presets (Table~\ref{tab:workloads}): \emph{small} for the main comparison and ablations, and \emph{long}, a simulated year and a half of sessions, for behaviour over time. Seeds 5--9 of the small preset are held out: no design decision was made on them.

\begin{table}[t]
\caption{Workloads. Synthetic columns give the range per seed (one timeline per seed); probes are summed over seeds. LongMemEval values are per history. Tokens: the text of all events of one seed or one history, in model tokens (Sec.~\ref{sec:protocol}).}
\label{tab:workloads}
\centering
\footnotesize
\setlength{\tabcolsep}{3pt}
\begin{tabular}{@{}lrrrr@{}}
\toprule
 & Small & Held-out & Long & LongMem- \\
 & s0--4 & s5--9 & s0--2 & Eval \\
\midrule
Users (histories) & 3 & 3 & 6 & 210 \\
Sessions & 48 & 48 & 360 & ${\approx}$48 \\
Events & 184--218 & 184--217 & 1{,}151--1{,}203 & ${\approx}$493 \\
Tokens & 1.8k--2.3k & 1.9k--2.4k & 11.3k--12.0k & 95k--104k \\
Facts & 22--34 & 20--33 & 60--81 & -- \\
Mutations & 63--105 & 64--100 & 242--329 & -- \\
Deletions & 4--7 & 5--9 & 20--32 & -- \\
Checkpoints & 4 & 4 & 8 & 1 \\
Probes (total) & 512 & 526 & 1{,}379 & 210 \\
\bottomrule
\end{tabular}
\end{table}

\subsection{LongMemEval}
\label{sec:lme}

LongMemEval\_S~\cite{wu2025longmemeval} has 500 questions, each after a
history of about 48 chat sessions. We group the questions into seven categories (six question types and abstention) and draw 30 from each with Python's \texttt{random.sample} and seed 0 (210 questions; ids in the artifact). Each history is replayed session by session in date order into a fresh store, and the question is asked at its date. Answers are judged with the benchmark's own prompts, verbatim, on its judge model \texttt{gpt-4o-2024-08-06} at temperature 0; the only change is that the judge returns \texttt{yes}/\texttt{no} in a JSON field.

Before the main run, a 14-question pilot, drawn in the same way, chose between the fallback and augment modes; we report both. Ten pilot questions are also in the 210, so mode selection saw part of the evaluation set and we do not treat it as untouched test data. Without them, accuracy changes by at most one point per system (augment 75.0\%, full replay 74.0\%), and no conclusion changes; the ten ids and these scores are in the artifact.

To check the judge, one author audited 60 of its decisions without knowing the system: for full replay, transcript RAG and \hyb{} in augment mode, 10 the judge accepted and 10 it rejected, covering all seven categories. The judge agreed with the human on 59 of 60 (98.3\%, Cohen's $\kappa=0.97$). Mem0, \sys{} and fallback mode were not audited, so this agreement does not validate every system's score.

\subsection{Protocol}
\label{sec:protocol}

\textbf{Systems.} No memory; full replay (the whole history, unbounded);
transcript RAG (BM25 over events, top 5, within the budget);
Mem0~\cite{chhikara2025mem0} (OSS release 1.0.11, unmodified, with a local vector store,
\texttt{text-embedding-3-small} and our extraction model; top 10), and the same
with its list of personal fact types replaced by our extractor's definition of
a coding fact, fixed before any run (\emph{coding prompt}); \sys{} with the LLM extractor and with a rule-based extractor (four hand-written regular expressions, a model-free floor, not a tuned baseline); and \hyb{}. Bounded systems share one evidence budget (2{,}000 tokens on the coding workload, 4{,}000 on LongMemEval); the runner truncates whole items, but in our runs it never cut any evidence (see below).

\textbf{Shared answerer.} Every system's evidence goes to the same model
(\texttt{gpt-4.1-mini-2025-04-14}, temperature 0, fixed seed) with the same
prompt, memory first and question last. The answer is structured (value,
abstain flag), so on the coding workload abstention is read from a field and
answers are scored deterministically against the planted truth with no LLM
judge. An empty or refused answer counts as an abstention.

\textbf{Two scores.} Each probe of the coding workload is scored twice. The \emph{evidence verdict} scores what memory sent to the model; the \emph{answer verdict} applies the same rules to the shared model's answer. A value counts as present when its text, or a fixed alias such as ``PG $N$'', appears as a whole phrase, ignoring case. Each forbidden value is labelled by why it is wrong: \emph{resurrected} (stated before the last deletion), and after it \emph{stale} (set before the target time), \emph{future} (set after the as-of time) or \emph{contaminated} (from a contradiction). In a historical question, the value valid at the as-of time is the correct answer, not a forbidden one.

Verdicts follow a fixed order. For a probe with an expected value: \emph{false abstention} if the system abstained; otherwise \emph{correct} (the value appears, no forbidden value does), \emph{polluted} (both appear), \emph{wrong} (only a forbidden value) or \emph{missed} (neither). For a probe whose truth is to abstain: \emph{leaked} if a forbidden value appears, even if the system also abstained; otherwise \emph{correct} if it abstained, or \emph{no abstention}. Deleted-value counts are read from the evidence, whatever the verdict. At the evidence level only an explicit abstention by the memory counts, so systems without an abstention signal (no memory, full replay, transcript RAG) score \emph{no abstention} on these probes. Only \emph{correct} counts as a success.

\textbf{Aggregation and intervals.} Evidence and answer accuracy are the share of correct verdicts over all probes of a seed. We report the mean and sample standard deviation over seeds; counts are pooled over seeds, and a probe counts once for each type of forbidden value it contains. On LongMemEval, each system runs once on the same 210 questions; accuracy describes the balanced subset, not the benchmark's own mix. Its 95\% interval is a percentile bootstrap over questions (2{,}000 resamples), and two systems are compared with a paired bootstrap on the same questions. We do not correct for multiple comparisons, so these comparisons are exploratory.

\textbf{Cost and efficiency.} The runner, not the system, measures latency, counts all injected tokens, and converts model usage to dollars with prices pinned on 2026-10-19 (input charged as uncached, an upper bound). All token counts are model tokens of the answering model's tokenizer (\texttt{o200k\_base}). Our runs counted words and punctuation marks when fitting the budget; the largest bounded evidence set was 205 model tokens on the coding workload and 2{,}538 on LongMemEval, so neither count cut any evidence, and we recount every logged evidence set with the model tokenizer. Latency is wall-clock time around each system call on one laptop (Apple M5, 16 GB), one call at a time, including model and embedding calls; LLM timings come from uncached repeats, because a cached replay is not a timing. We report write cost per session or history, read and answering cost per question, injected tokens, latency and store size.

\textbf{Reproducibility.} Extraction, routing, Mem0's writes and answering use \texttt{gpt-4.1-mini}; every logged response reports the snapshot \texttt{gpt-4.1-mini-2025-04-14}. Sampling is pinned (temperature 0 and a fixed seed); every response is cached on disk under a hash of provider, model, prompt and schema, so a run replays at zero cost. Five seeds give variance across workloads; repeated runs with the cache bypassed measure the model's own noise. Code, workloads, pinned prices, all run logs, sampled ids, per-seed results, the pilot analysis, the cost breakdown and the supplement are available at
\url{https://anonymous.4open.science/r/memory-challenge-4482/}.

\subsection{Coverage of the challenge requirements}
\label{sec:coverage}

We measure directly persistence across sessions and restarts, provenance, update, supersession and deletion, bounded evidence and abstention (R1); an established benchmark, required baselines under the same model, prompt and budget, and quality, tokens, cost, ingest time, latency and storage (R2); and checkpoints, typed stale errors, growth, pinned models and repeated runs (R3). We cover four only in part: our traces are heterogeneous but generated from templates; latency growth over time is not reported, because the long-preset replays were cached; the judge audit covers three of seven configurations; and, the main gap, no agent performs a downstream task. A shared model answers questions at checkpoints, so we measure whether memory provides the right facts, not whether an agent fixes a bug or deploys a service better with memory. The full mapping is in the supplement (S2).
\section{Results}
\label{sec:results}

\begin{table*}[t]
\caption{Coding-agent workload, small preset, seeds 0--4 (512 probes). Answer and evidence accuracy (\%, mean $\pm$ sd over seeds); counts pooled over seeds (Sec.~\ref{sec:protocol}). Forbidden in evidence: number of probes whose evidence held at least one stale, future or contaminated value, then the probes per type; types overlap, so they do not sum to the first number. Deleted value: evidence/answer, deletion requests with description only. Tokens: mean injected per probe, in model tokens. \$: memory plus answering, all seeds.}
\label{tab:coding}
\centering
\footnotesize
\begin{tabular}{@{}lccrcccr@{}}
\toprule
System & Answer & Evidence & Wrong ans. & Forbidden in evidence: probes (s/f/c) & Deleted value & Tokens & \$ \\
\midrule
No memory & 29.5 $\pm$ 1.5 & 0.0 $\pm$ 0.0 & 0 & 0 (0/0/0) & 0 / 0 & 0 & 0.05 \\
Full replay & 82.1 $\pm$ 3.0 & 21.9 $\pm$ 1.8 & 65 & 249 (129/93/93) & 46 / 4 & 1{,}069 & 0.28 \\
Transcript RAG & 74.8 $\pm$ 3.6 & 27.0 $\pm$ 5.0 & 88 & 195 (101/69/66) & 39 / 0 & 128 & 0.08 \\
Mem0 & 66.8 $\pm$ 1.8 & 34.9 $\pm$ 2.7 & 66 & 107 (45/56/13) & 18 / 17 & 141 & 0.37 \\
Mem0 (coding prompt) & 70.6 $\pm$ 6.4 & 34.7 $\pm$ 2.9 & 82 & 143 (44/76/36) & 16 / 13 & 158 & 0.43 \\
\sys{} (rules) & 39.5 $\pm$ 6.9 & 38.7 $\pm$ 6.3 & 9 & 9 (9/0/0) & 0 / 0 & 5 & 0.05 \\
\sys{} (LLM) & 90.9 $\pm$ 5.5 & 91.7 $\pm$ 5.2 & 4 & 4 (4/0/1) & 0 / 0 & 17 & 0.23 \\
\hyb{} fallback & 94.3 $\pm$ 4.2 & 66.8 $\pm$ 4.6 & 10 & 16 (6/8/3) & 0 / 0 & 61 & 0.23 \\
\hyb{} augment & 94.4 $\pm$ 2.2 & 32.4 $\pm$ 7.8 & 15 & 193 (104/70/60) & 0 / 0 & 144 & 0.25 \\
\bottomrule
\end{tabular}
\end{table*}

\subsection{Wrong values in the evidence (RQ1)}
\label{sec:rq1}

\emph{Trust.} With the whole history in context, full replay's evidence is
clean in only 21.9\% of probes (Table~\ref{tab:coding}). Its answers are right in 82.1\% of probes,
but 65 of them contain a forbidden value. In 45 of these the model sides with
a stale document or with the assistant's own guess over the user, and it
answers only 31 of 67 contamination probes. \sys{} answers 62 of 67, because
the less trusted claim never reaches the context. Its evidence is correct in
91.7\% of probes. Four of its answers contain a forbidden value, all stale values from updates the extractor missed.

\emph{Model-decided updates.} Mem0 lets the model add,
update or delete each extracted fact. An update overwrites the memory in place, so the
value an earlier date needs is gone: Mem0 answers 27 of 93 historical
questions (\sys{} 69, full replay 74), and 27 of its answers give a later
value for an earlier date. With the coding prompt it answers more current questions (96 instead of 86 of 112), but its
contaminated answers rise from 10 to 27. Mem0 and \sys{} also differ in extraction, routing and retrieval, so this comparison does not isolate one cause.

\emph{Deletion.} When deletion requests carry only a description, the transcript baselines remove nothing. Full replay sends the deleted value to the model in 46 probes and repeats it to the user in 4; transcript RAG sends it in 39. At answer level the leak is mostly hidden, because the model also sees the ``forget it'' message. For Mem0 the request is handled only by its ordinary update step, which kept the value: it reached the model in 18 probes and the user in 17 (16 and 13 with the coding prompt). This is a result for the evaluated configuration. \sys{} and \hyb{} sent no deleted value, in either mode.

\subsection{Answer accuracy (RQ2)}
\label{sec:rq2}

\emph{Coding workload, development seeds (0--4).} We developed the rules and
safeguards on these seeds. \hyb{} answers 94.3\% (fallback) and 94.4\%
(augment), \sys{} 90.9\%, full replay 82.1\%, transcript RAG 74.8\%, and
Mem0 66.8\% (70.6\% with the coding prompt). \hyb{} gains over \sys{} mostly
by answering historical questions on which the ledger abstained. In fallback mode, 16 probes have at least one forbidden value in the evidence, against 249 for full replay and 195 for transcript RAG. In augment mode there are 193, so its correctness depends on the answering model preferring the ledger's fact line, which comes first.

\begin{table}[t]
\caption{Held-out seeds 5--9 of the small preset (526 probes), never inspected
during development: answer and evidence accuracy (\%, mean $\pm$ sd over seeds)
and answers containing a forbidden value.}
\label{tab:heldout}
\centering
\footnotesize
\begin{tabular}{@{}lccr@{}}
\toprule
System & Answer & Evidence & Wrong ans. \\
\midrule
No memory & 29.4 $\pm$ 2.9 & 0.0 $\pm$ 0.0 & 0 \\
Full replay & 83.3 $\pm$ 2.8 & 21.1 $\pm$ 0.7 & 58 \\
Transcript RAG & 78.5 $\pm$ 3.1 & 23.8 $\pm$ 3.6 & 81 \\
Mem0 & 66.2 $\pm$ 7.1 & 34.8 $\pm$ 6.2 & 59 \\
Mem0 (coding prompt) & 68.4 $\pm$ 4.3 & 36.0 $\pm$ 3.1 & 94 \\
\sys{} (rules) & 37.0 $\pm$ 2.2 & 37.0 $\pm$ 2.2 & 13 \\
\sys{} (LLM) & 88.2 $\pm$ 3.2 & 89.9 $\pm$ 1.7 & 5 \\
\hyb{} fallback & 91.1 $\pm$ 5.4 & 64.2 $\pm$ 4.8 & 15 \\
\hyb{} augment & 91.8 $\pm$ 4.1 & 25.9 $\pm$ 3.8 & 23 \\
\bottomrule
\end{tabular}
\end{table}

\emph{Coding workload, held-out seeds (5--9).} We ran these seeds once, with the
final systems (Table~\ref{tab:heldout}). \hyb{} in fallback mode drops from 94.3\% to 91.1\%, while full replay rises from 82.1\% to 83.3\%, so its lead shrinks from 12.2 to 7.8 points. In augment mode the lead shrinks from 12.3 to 8.5 points, and for \sys{} from 8.8 to 4.9. The development seeds therefore flatter our systems. The ranking of the systems holds.

\begin{table*}[t]
\caption{LongMemEval\_S, stratified subset of 210 questions (30 per category).
Accuracy (\%) with a 95\% bootstrap interval, by category, and mean injected model tokens. Costs in milli-dollars: Write, to ingest one history, paid once; Read (retrieval, including routing) and Answer, paid for every question. Each question has its own history here; when several questions share a history, only the write cost is divided among them. Judge: \texttt{gpt-4o-2024-08-06} with the benchmark's prompts.}
\label{tab:lme}
\centering
\footnotesize
\begin{tabular}{@{}lc ccccccc rrrr@{}}
\toprule
System & Accuracy & SS-U & SS-A & Pref & Multi & KU & Temp & Abs & Tokens & Write & Read & Ans. \\
\midrule
No memory & 16.2 (11--21) & 0 & 0 & 13 & 0 & 0 & 0 & 100 & 0 & 0.0 & 0.0 & 0.1 \\
Full replay & 74.8 (69--80) & 100 & 97 & 63 & 53 & 90 & 60 & 60 & 108{,}447 & 0.0 & 0.0 & 43.5 \\
Transcript RAG & 71.4 (65--78) & 100 & 97 & 33 & 43 & 73 & 67 & 87 & 1{,}072 & 0.0 & 0.0 & 0.6 \\
Mem0 & 71.0 (65--77) & 83 & 40 & 70 & 53 & 83 & 80 & 87 & 221 & 160.9 & 0.0 & 0.2 \\
\sys{} (LLM) & 41.0 (34--48) & 63 & 3 & 13 & 40 & 60 & 13 & 93 & 15 & 49.5 & 1.5 & 0.1 \\
\hyb{} fallback & 70.5 (65--76) & 97 & 93 & 30 & 50 & 67 & 67 & 90 & 593 & 49.5 & 1.5 & 0.4 \\
\hyb{} augment & 74.8 (69--80) & 97 & 97 & 33 & 60 & 87 & 67 & 83 & 1{,}087 & 49.5 & 1.5 & 0.6 \\
\bottomrule
\end{tabular}
\end{table*}

\emph{LongMemEval.} Full replay, now one 108k-token context per question, reaches 74.8\% (Table~\ref{tab:lme}). \hyb{} (augment) has the same observed accuracy, 74.8\%, with 1{,}087 injected tokens, 1\% of the context. The paired 95\% interval of the difference ($-6.2$ to $+6.2$ points) still allows a real difference of several points either way, so this is not evidence of equivalence. Against transcript RAG it gains 3.3 points (CI 0.0 to $+7.1$), mainly on knowledge-update (87 vs.\ 73) and multi-session questions (60 vs.\ 43). Mem0 reaches 71.0\% with 221 tokens; \hyb{} (augment) is 3.8 points ahead (CI $-3.8$ to $+11.4$), so this sample does not establish a difference. Mem0 leads on preference (70) and temporal questions (80), the personal-chat setting its prompt was written for, but its write cost is three times ours (161 vs.\ 50 milli-dollars per history). The ledger alone answers only 41.0\%: the benchmark mostly asks for one detail buried in unrelated chats, and by design the ledger never answers from the assistant's statements (SS-A 3\%).

\subsection{Growth over time (RQ3)}
\label{sec:rq3}

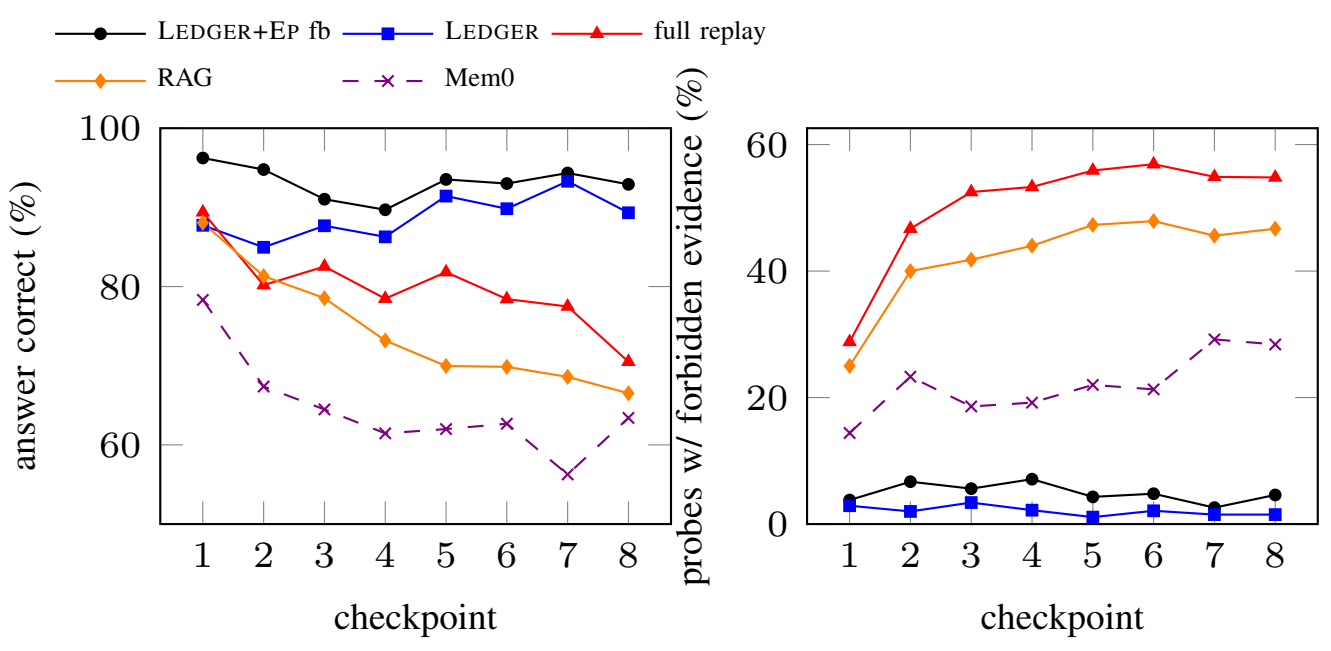
\begin{figure}[t]
\centering
\begin{tikzpicture}
\begin{axis}[name=acc, width=0.56\columnwidth, height=4.2cm, xlabel={checkpoint}, ylabel={answer correct (\%)},
  ymin=50, ymax=100, xtick={1,...,8}, tick label style={font=\scriptsize}, label style={font=\scriptsize},
  legend style={font=\tiny, at={(0.5,1.03)}, anchor=south, draw=none, fill=none}, legend columns=3, legend cell align=left]
\addplot[black, mark=*, mark size=1pt] table[x=cp, y=answer, row sep=\\] {cp answer evidence forbidden kib tokens retrievems \\ 1 96.24 59.46 3.8 5.91 68 0.784 \\ 2 94.78 64.74 6.7 11.64 60 0.545 \\ 3 91.03 65.53 5.6 17.07 59 0.574 \\ 4 89.70 63.36 7.1 22.07 61 0.455 \\ 5 93.53 68.86 4.3 27.04 56 0.486 \\ 6 93.01 66.54 4.8 32.18 57 0.558 \\ 7 94.34 66.15 2.6 36.43 58 0.512 \\ 8 92.91 65.99 4.6 41.06 59 0.487 \\};
\addplot[blue, mark=square*, mark size=1pt] table[x=cp, y=answer, row sep=\\] {cp answer evidence forbidden kib tokens retrievems \\ 1 87.74 87.89 2.9 0.32 14 640.306 \\ 2 84.96 86.94 2.0 0.55 16 656.678 \\ 3 87.68 88.22 3.4 0.73 17 638.114 \\ 4 86.28 88.42 2.2 0.84 17 637.343 \\ 5 91.43 90.38 1.1 1.01 18 655.138 \\ 6 89.83 89.83 2.1 1.14 18 670.261 \\ 7 93.30 93.79 1.5 1.19 18 636.163 \\ 8 89.32 88.81 1.5 1.28 17 631.447 \\};
\addplot[red, mark=triangle*, mark size=1.2pt] table[x=cp, y=answer, row sep=\\] {cp answer evidence forbidden kib tokens retrievems \\ 1 89.38 33.87 28.8 5.66 598 0.350 \\ 2 80.18 23.46 46.7 11.42 1154 0.659 \\ 3 82.53 19.77 52.5 16.94 1716 0.982 \\ 4 78.47 18.75 53.3 22.40 2268 1.205 \\ 5 81.82 17.78 55.9 27.70 2807 1.752 \\ 6 78.42 15.99 56.9 33.23 3355 1.680 \\ 7 77.48 16.92 54.9 38.10 3868 2.517 \\ 8 70.50 16.74 54.8 43.46 4399 1.705 \\};
\addplot[orange, mark=diamond*, mark size=1.2pt] table[x=cp, y=answer, row sep=\\] {cp answer evidence forbidden kib tokens retrievems \\ 1 88.05 35.76 25.0 5.66 127 0.445 \\ 2 81.32 27.29 40.0 11.42 128 0.658 \\ 3 78.53 24.84 41.8 16.94 129 0.892 \\ 4 73.18 22.65 44.0 22.40 129 1.088 \\ 5 69.97 18.93 47.3 27.70 130 1.478 \\ 6 69.86 19.66 47.9 33.23 131 1.758 \\ 7 68.59 17.43 45.6 38.10 130 1.649 \\ 8 66.52 14.72 46.7 43.46 131 2.336 \\};
\addplot[violet, dashed, mark=x, mark size=1.5pt, mark options={solid}] table[x=cp, y=answer, row sep=\\] {cp answer evidence forbidden kib tokens retrievems \\ 1 78.32 40.38 14.4 1.74 110 52.336 \\ 2 67.37 37.21 23.3 2.97 161 67.940 \\ 3 64.48 36.75 18.6 3.75 175 67.224 \\ 4 61.46 32.48 19.2 4.33 187 66.300 \\ 5 62.00 34.17 22.0 4.84 187 65.203 \\ 6 62.68 36.84 21.3 5.59 186 74.735 \\ 7 56.27 23.43 29.2 6.03 185 53.739 \\ 8 63.41 29.42 28.4 6.47 187 59.519 \\};
\legend{\hyb{} fb, \sys{}, full replay, RAG, Mem0}
\end{axis}
\begin{axis}[at={(acc.east)}, anchor=west, xshift=0.9cm, width=0.56\columnwidth, height=4.2cm, xlabel={checkpoint},
  ylabel={probes w/ forbidden evidence (\%)}, ymin=0, xtick={1,...,8}, tick label style={font=\scriptsize}, label style={font=\scriptsize}]
\addplot[black, mark=*, mark size=1pt] table[x=cp, y=forbidden, row sep=\\] {cp answer evidence forbidden kib tokens retrievems \\ 1 96.24 59.46 3.8 5.91 68 0.784 \\ 2 94.78 64.74 6.7 11.64 60 0.545 \\ 3 91.03 65.53 5.6 17.07 59 0.574 \\ 4 89.70 63.36 7.1 22.07 61 0.455 \\ 5 93.53 68.86 4.3 27.04 56 0.486 \\ 6 93.01 66.54 4.8 32.18 57 0.558 \\ 7 94.34 66.15 2.6 36.43 58 0.512 \\ 8 92.91 65.99 4.6 41.06 59 0.487 \\};
\addplot[blue, mark=square*, mark size=1pt] table[x=cp, y=forbidden, row sep=\\] {cp answer evidence forbidden kib tokens retrievems \\ 1 87.74 87.89 2.9 0.32 14 640.306 \\ 2 84.96 86.94 2.0 0.55 16 656.678 \\ 3 87.68 88.22 3.4 0.73 17 638.114 \\ 4 86.28 88.42 2.2 0.84 17 637.343 \\ 5 91.43 90.38 1.1 1.01 18 655.138 \\ 6 89.83 89.83 2.1 1.14 18 670.261 \\ 7 93.30 93.79 1.5 1.19 18 636.163 \\ 8 89.32 88.81 1.5 1.28 17 631.447 \\};
\addplot[red, mark=triangle*, mark size=1.2pt] table[x=cp, y=forbidden, row sep=\\] {cp answer evidence forbidden kib tokens retrievems \\ 1 89.38 33.87 28.8 5.66 598 0.350 \\ 2 80.18 23.46 46.7 11.42 1154 0.659 \\ 3 82.53 19.77 52.5 16.94 1716 0.982 \\ 4 78.47 18.75 53.3 22.40 2268 1.205 \\ 5 81.82 17.78 55.9 27.70 2807 1.752 \\ 6 78.42 15.99 56.9 33.23 3355 1.680 \\ 7 77.48 16.92 54.9 38.10 3868 2.517 \\ 8 70.50 16.74 54.8 43.46 4399 1.705 \\};
\addplot[orange, mark=diamond*, mark size=1.2pt] table[x=cp, y=forbidden, row sep=\\] {cp answer evidence forbidden kib tokens retrievems \\ 1 88.05 35.76 25.0 5.66 127 0.445 \\ 2 81.32 27.29 40.0 11.42 128 0.658 \\ 3 78.53 24.84 41.8 16.94 129 0.892 \\ 4 73.18 22.65 44.0 22.40 129 1.088 \\ 5 69.97 18.93 47.3 27.70 130 1.478 \\ 6 69.86 19.66 47.9 33.23 131 1.758 \\ 7 68.59 17.43 45.6 38.10 130 1.649 \\ 8 66.52 14.72 46.7 43.46 131 2.336 \\};
\addplot[violet, dashed, mark=x, mark size=1.5pt, mark options={solid}] table[x=cp, y=forbidden, row sep=\\] {cp answer evidence forbidden kib tokens retrievems \\ 1 78.32 40.38 14.4 1.74 110 52.336 \\ 2 67.37 37.21 23.3 2.97 161 67.940 \\ 3 64.48 36.75 18.6 3.75 175 67.224 \\ 4 61.46 32.48 19.2 4.33 187 66.300 \\ 5 62.00 34.17 22.0 4.84 187 65.203 \\ 6 62.68 36.84 21.3 5.59 186 74.735 \\ 7 56.27 23.43 29.2 6.03 185 53.739 \\ 8 63.41 29.42 28.4 6.47 187 59.519 \\};
\end{axis}
\end{tikzpicture}
\caption{Long preset (6 users, 60 sessions each over 540 simulated days, 8 checkpoints; mean of 3 seeds). Left: answer accuracy per checkpoint. Right: share of probes per checkpoint whose evidence carried a stale, future or contaminated value (104--197 probes per checkpoint).}
\label{fig:long}
\end{figure}

Fig.~\ref{fig:long} replays 540 simulated days on the long preset, with
eight checkpoints. As the history grows from 598 to 4{,}399 tokens per query,
full replay falls from 89.4\% to 70.5\% and transcript RAG from 88.0\% to
66.5\%, and the share of their probes with a forbidden value in the evidence grows from 29\% to 55\% and from 25\% to 47\%. \sys{} stays between 85.0\% and 93.3\%, with 1--3\% of probes affected, and \hyb{} (fallback) between 89.7\% and 96.2\%, with 3--7\%. Mem0 starts at 78.3\% and falls to 56--63\% from the fourth checkpoint on. Its store stays small (6.5\,KiB), so
it does not drown in history; it fails on time instead: 312 of its 1{,}379 probes carry a forbidden value in the evidence, in 170 a future value and in 130 a stale one.

\emph{Endpoint cost.} At the last checkpoint, transcript RAG injects about 130 tokens per question, \sys{} 17 and \hyb{} 59, against 4{,}399 for full replay. The ledger stores 1.3\,KiB per seed against 43.5\,KiB for the transcript store; \hyb{} keeps both in 41.1\,KiB, because its deletion scrubbing removes 43--80 more turns per seed. Our systems make at most one extraction call per session and one routing call per question throughout, but each call resends the list of known slots: extraction input grows from 418 to 572 tokens per session and routing input from 319 to 508 tokens per question.

\subsection{Operational reliability}
\label{sec:ops}

\textbf{Restarts.} We tested five systems: full replay, transcript RAG, \sys{} with both extractors, and \hyb{} in augment mode. After every checkpoint of the long preset, we wrote each store to disk, restored it into a fresh instance and continued. For every system, all 1{,}379 probes got the same verdict and evidence as in the run without restarts. Snapshots stay under 270\,KiB and restore in at most 63\,ms.

\textbf{Model change.} In \hyb{} (augment mode, long preset, seeds 0 and 1), we switched the model behind extraction and routing from \texttt{gpt-4.1-mini} to \texttt{gpt-4.1} after the fourth of eight checkpoints and kept the store. No migration was needed. Second-half accuracy rises from 92.0\% (mini throughout) to 95.4\%, within 0.3 points of a run on \texttt{gpt-4.1} throughout (95.7\%), at \$1.18 for extraction and routing against \$1.89 on \texttt{gpt-4.1} throughout.

\textbf{Run-to-run noise.} We ran \sys{} and \hyb{} (fallback) three times per seed, twice with the response cache bypassed. In 12 and 8 of the 512 probes, the answer verdict differed in at least one run. The standard deviation of accuracy across repeats is 0.7 and 0.4 points per seed, far below the differences between systems.

\subsection{Overall cost and efficiency}
\label{sec:efficiency}

\begin{table*}[t]
\caption{Efficiency on the coding workload (small preset, seeds 0--4):
write-time LLM calls per ingested session, median ingest time per session and
retrieval time per query, store size at the end per seed, mean injected model tokens
per probe, and cost in milli-dollars to write, per ingested session, and to read and answer, per probe.
Times of the LLM systems are from uncached repeats (cached replays are not
timings); augment mode shares fallback's write and routing path. Mem0 with the
coding prompt is timed from an uncached rerun, since its logged run reused Mem0's cached embeddings.}
\label{tab:efficiency}
\centering
\footnotesize
\begin{tabular}{@{}lrrrrrrrr@{}}
\toprule
System & LLM calls/session & Ingest ms & Retrieve ms & Store KiB & Tokens & Write m\$ & Read m\$ & Ans m\$ \\
\midrule
No memory & 0.00 & 0.00 & 0.00 & 0.0 & 0 & 0.00 & 0.00 & 0.10 \\
Full replay & 0.00 & 0.00 & 0.16 & 8.1 & 1{,}069 & 0.00 & 0.00 & 0.55 \\
Transcript RAG & 0.00 & 0.00 & 0.35 & 8.1 & 128 & 0.00 & 0.00 & 0.16 \\
Mem0 & 1.65 & 2989 & 174 & 1.5 & 141 & 1.06 & 0.00 & 0.16 \\
Mem0 (coding prompt) & 1.83 & 3272 & 202 & 1.9 & 158 & 1.24 & 0.00 & 0.17 \\
\sys{} (rules) & 0.00 & 0.03 & 0.02 & 0.1 & 5 & 0.00 & 0.00 & 0.11 \\
\sys{} (LLM) & 1.00 & 836 & 666 & 0.4 & 17 & 0.31 & 0.18 & 0.11 \\
\hyb{} fallback & 0.93 & 817 & 662 & 8.0 & 61 & 0.27 & 0.17 & 0.13 \\
\hyb{} augment & 0.93 & 817 & 662 & 8.0 & 144 & 0.27 & 0.17 & 0.16 \\
\bottomrule
\end{tabular}
\end{table*}

Table~\ref{tab:efficiency} gives the cost of the write and read paths on the small preset. The transcript systems retrieve in under a millisecond, but inject more context: 1{,}069 tokens per probe for full replay and 128 for transcript RAG, against 17 for the ledger and 61 for \hyb{} (fallback). Reading and answering cost \hyb{} (fallback) 0.30\,m\$ per probe against 0.55\,m\$ for full replay, plus 0.27\,m\$ per session to write; over all seeds, \$0.23 against \$0.28.

\subsection{What still fails}
\label{sec:fails}

On the coding workload \sys{} fails 47 of 512 probes. Twenty are routing
misses (``no known slot matches''), mostly paraphrases the router does not
link to a slot (``where does it publish its events'' for a message-queue
slot) or slots filed under a short service name. Sixteen are historical questions
asked before the first value the extractor caught: when a fact's first
statement is terse or indirect, every earlier as-of question abstains. Four are stale answers from
updates stated indirectly (``build \#812 reported by Jenkins''), the only probes in which a forbidden value reaches \sys{}'s evidence; two are assistant guesses
correctly held but leaving the question unanswered; and in five the evidence
was right but the answering model abstained. None is a rule applied wrongly.

\hyb{} (fallback) fails 29. Ten answers carry a forbidden value: four are the ledger's own stale facts above, and six are old values taken from a raw turn. Four answers come where the truth is to abstain, none with a deleted value. In the other 15 the answering model abstained, mostly because BM25 did not find the right turn. On LongMemEval the dominant failure is recall of a single buried detail: \texttt{gpt-4.1-mini} extracts many incidental ``facts'' per task chat and misses
some personal details, and preference questions (33\% for \hyb{}) need a summary that neither facts nor five turns give.

\subsection{Ablations}
\label{sec:ablation}

\begin{table}[t]
\caption{Ablations on the coding workload (small preset, seeds 0--4). Development-time safeguard and prompt changes are in the supplement (S4). Evidence and answer: \% correct;
other metrics: number of probes.}
\label{tab:ablation}
\centering
\footnotesize
\setlength{\tabcolsep}{3pt}
\begin{tabular}{@{}llrr@{}}
\toprule
Change & Metric & Before & After \\
\midrule
\multicolumn{4}{@{}l}{\emph{(a) Single-factor trust-rule changes (cached)}} \\
Hold assistant-first values (trust 2)$^{*}$ & wrong answers & 6.0 & 4.0 \\
Hold document-first values too (trust 3)$^{*}$ & evidence & 91.5 & 89.7 \\
\multicolumn{4}{@{}l}{\emph{(a$'$) Complete-configuration comparisons}} \\
Rule-based $\to$ LLM extractor & evidence & 38.7 & 91.7 \\
\sys{} $\to$ \hyb{} fallback & answer & 90.9 & 94.3 \\
\bottomrule
\multicolumn{4}{@{}l}{\scriptsize $^{*}$Replayed from cache: every extraction identical.}
\end{tabular}
\end{table}

\emph{(a) Single-factor trust-rule changes} change one factor and keep everything else.
Both rows replay every extraction from cache, so only the rule
changes and the difference is exact. Holding the assistant's first values
(trust 2) turns two of six wrong answers into abstentions and changes nothing
else. Holding documents as well (trust 3) costs nine correct answers and
fixes none, so documents may still introduce a value.

\emph{(a$'$) Complete-configuration comparisons} come from the main runs (Table~\ref{tab:coding}): same seeds, answerer and budget. They change more than one component. The extractor row changes the whole extraction method. The \sys{} $\to$ \hyb{} row adds the episode store but also stops extracting low-trust events, so it does not isolate the episode store. Their differences exceed the run-to-run noise, but the same seeds do not remove this confounding.

The safeguards of Sec.~\ref{sec:extraction} and the prompt revisions were each measured once during development, before and after the change; these single runs are in the supplement (S4) and establish no stable, isolated effect. We did not run leave-one-out ablations of the final system, or ablate the trust ranking itself, the episode retriever or the token budget.

\subsection{Limits of the results}
\label{sec:limits}

The coding workload is synthetic, built by us, and phrased by templates, and our lead over full replay is about four points smaller on held-out seeds. Its histories stay short (about 4.4k tokens per user at the end), so the long preset tests growth in facts and updates more than in context length. The runs include extraction and routing errors, so they show how close the full system comes to the guarantees of Sec.~\ref{sec:problem}, not that they hold. Restarts happen only between checkpoints, and the model change covers one upgrade within one model family, on two seeds.
\section{Discussion}
\label{sec:discussion}

\textbf{The evidence--answer gap.} The answering model hides much of the difference between memories. It often picks the right value from noisy evidence, so full replay answers 82.1\% of the coding probes while its evidence is clean in only 21.9\%. This repair weakens as history grows, and it never helps privacy: a deleted value that the model does not repeat has still been sent to the provider. In our runs, the largest differences between memories came from deciding what is current, which source wins and what was deleted, more than from finding facts. Memory should therefore be judged on its evidence as well as on its answers.

\textbf{Configuration trade-offs.} The variants differ in who is responsible for correctness. The ledger alone sends only values that passed its rules, but abstains when it has no fact. Fallback mode adds raw turns only on those questions, so it keeps forbidden evidence low (16 probes against 193 for augment) at almost the same accuracy; we recommend it where facts change and sources disagree. Augment mode relies on the answering model to prefer the fact line and suits long chats with many one-off details, as on LongMemEval. Both move cost from reading to writing: with $c_w$ the write cost per history, $c_r$ and $c_a$ the read and answering costs per question, and $c_f$ full replay's cost per question, \hyb{} is cheaper after $k > c_w/(c_f - c_r - c_a)$ questions per history, on LongMemEval the second question ($c_w=49.5$, $c_r=1.5$, $c_a=0.6$, $c_f=43.5$\,m\$). The write cost grows with the number of known slots, not with the length of the raw history.
\section{Threats to Validity}
\label{sec:threats}

\textbf{Construct validity.} We built both the coding workload and the system, so the workload may favour our design. Every system runs on the same timelines with the same answerer, and we also report LongMemEval, which we did not design. Our workload states every change when it happens, so it never tests A4, and how often real traffic breaks A3 or A4 is untested. The LongMemEval judge is a model, audited for three of the seven configurations.

\textbf{Internal validity.} We developed the rules and safeguards while looking at failures on seeds 0--4, and we report results on these seeds. Held-out seeds 5--9, run once, show that the results do not depend on the timelines we inspected, but not that they hold outside the generator, because new seeds use the same templates, fact kinds and mutation rules. The LongMemEval mode choice used a pilot that overlaps the evaluation set.

\textbf{External validity.} The coding workload uses templated phrasings and nine kinds of facts. Real traces are messier, and other fact kinds or phrasings may be harder to extract and route. All extraction, routing and answering use one model family (\texttt{gpt-4.1-mini}, and \texttt{gpt-4.1} in the model-change run), and all text is English. Other model families may extract differently. Besides the baselines, we compare against one memory product, Mem0; Zep and A-MEM are not evaluated.

\textbf{Conclusion validity.} The coding results use five seeds, and the standard deviation of our systems is up to 5.5 points. The LongMemEval subset is 210 of 500 questions. Differences of a few points, such as between \hyb{} and Mem0 or full replay, lie within the paired confidence intervals, which are not corrected for multiple comparisons; small differences should not be read as rankings.
\section{Related Work}
\label{sec:related}

\textbf{Persistent agent memory.} Several systems give agents memory that lasts beyond one context window. MemGPT~\cite{packer2023memgpt} moves text between the context and external storage, and the model edits its own memory through function calls. Mem0~\cite{chhikara2025mem0} extracts facts from each session and lets the model add, update or delete them. A-MEM~\cite{xu2025amem} stores notes that the model links and rewrites over time. Zep~\cite{rasmussen2025zep} keeps a temporal knowledge graph whose edges record when a fact held, and uses the model to find contradicting edges. Other work stores reflections~\cite{park2023generative}, reusable skills and workflows~\cite{wang2023voyager,wang2024awm}, or a graph index for retrieval~\cite{gutierrez2024hipporag}. In all of these systems the model decides when a stored fact is replaced or removed; in \sys{} it only proposes claims, and fixed rules decide. Zep is closest to our design, since it also keeps when a fact held, and it also tries to read such dates from the text, which our ledger does not. Mem0 is closest to our pipeline, but an update overwrites the stored text, so the value an earlier date needs is gone.

\textbf{Temporal and provenance-aware data management.} The ideas we use for storage are well established. Temporal databases keep the time during which a fact was true (valid time) apart from the time it was recorded, so a change can be entered after it happened, and they close an interval instead of overwriting a row~\cite{snodgrass1999temporal}. Our ledger uses only a restricted form: valid time starts when a value is stated (A4), so it does not recover the real date of a change reported later. Provenance research records which source produced each piece of data~\cite{buneman2001provenance}. Our contribution is to bring them into agent memory, where the facts come from a language model and not from a trusted application. This requires parts that a database does not have: a trust order over sources that decides which value wins, a lexical check that rejects claims whose value does not appear in their source event, and deletion that also removes matching raw turns and blocks their source events from being written again.

\textbf{Memory evaluation.} LoCoMo~\cite{maharana2024locomo} tests recall over very long conversations. LongMemEval~\cite{wu2025longmemeval} tests five abilities, including knowledge updates, temporal reasoning and abstention. MemoryAgentBench~\cite{hu2025memoryagentbench} adds conflict resolution and finds that no current method masters all of its competencies. These benchmarks score the final answer, and their histories contain no deletion requests. Our evaluation also scores the evidence, labels each wrong value by why it is wrong, and tests deletion by description, restarts and a model change.
\section{Conclusion}

We presented \sys{}, an agent memory in which a model proposes facts and explicit rules decide which value is current, which source wins and what is deleted, and \hyb{}, which adds raw turns. Our strongest supported finding is about evidence: on a coding workload with planted changes, conflicting sources and deletion requests, \hyb{} in fallback mode sent a forbidden value in 16 of 512 probes, against 249 for full replay, and no deleted value, while answering more questions correctly on both development and held-out seeds. On LongMemEval it reached the same observed accuracy as full replay with 1\% of the context. The main limitation is that the coding workload is synthetic and built by us, and that we measure answers to questions, not an agent's success on downstream tasks.

\bibliographystyle{IEEEtran}
\bibliography{refs}

\end{document}